\documentstyle[epsf,floats,preprint,aps]{revtex}
\renewcommand\vec[1]{{\bf #1}}
\newcommand\carbon{{}^{12}{\rm C}}
\newcommand\argon{{}^{40}\!{\rm Ar}}
\title{
Improvement of the nucleon emission process and the statistical
property in molecular dynamics}
\author{Akira Ono}
\address{Institute of Physical and Chemical Research (RIKEN), Wako,
Saitama 351-01, Japan}
\author{Hisashi Horiuchi}
\address{Department of Physics, Kyoto University, Kyoto 606-01, Japan}
\draft
\begin{document}
\maketitle
\begin{abstract}
We propose to introduce a new stochastic process in molecular dynamics
in order to improve the description of the nucleon emission process
from a hot nucleus. We give momentum fluctuations originating from the
momentum width of the nucleon wave packet to the nucleon
stochastically when it is being emitted from the nucleus.  We show by
calculating the liquid gas phase equilibrium in the case of
antisymmetrized molecular dynamics, that with this improvement, we can
recover the quantum mechanical statistical property of the nucleus for
the particle emission process.
\end{abstract}

\narrowtext
In multifragmentation reactions in intermediate energy heavy ion
collisions, many intermediate mass fragments are produced. Data of the
mass distribution and other exclusive quantities have been reproduced
rather well by the statistical models \cite{BONDORF,GROSS} which
assume the decay of the thermal source into fragments according to the
statistical weights of the possible final states.  It is therefore
reasonably considered that the multifragmentation process is largely
governed by the statistical property. On the other hand, molecular
dynamics simulations try to describe the fragmentation based on the
microscopic dynamical evolution of the system
\cite{AICHELIN,BOAL,FELDMEIER}.
With molecular dynamics simulations we can study the dynamical and
statistical aspects in a unified framework without assuming any
thermal equilibrium. However, for the reliable description of the
fragmentation, we should check the validity of the statistical aspect
of molecular dynamics and, if necessary, modify the dynamics of
molecular dynamics so as to get the appropriate statistical aspect.

In our previous studies \cite{ONOabc,ONOde} of heavy ion collisions
with antisymmetrized molecular dynamics (AMD) in the incident energy
region between 30 MeV/nucleon and 100 MeV/nucleon, we observed that
light nuclei such as $\carbon$ break up into fragments easily in the
dynamical stage of the reaction but intermediate mass fragments are
seldom produced in the collision of heavier nuclei such as
$\argon$. We are also aware that the decay of excited nuclei produced
in the reaction is too slow when the AMD calculation is continued for
a long time, compared to the prediction by the statistical decay
model.  These points strongly suggested that we should improve
particle emission processes in AMD. In this letter, we propose a
method how to improve AMD in the nucleon emission process from the
excited nucleus and show by calculating the liquid gas phase
equilibrium that the improved AMD describes the quantum mechanical
statistical property of hot nuclei for the particle emission process
while the original AMD describes only the classical statistical
property for that process.  Our method of improvement is of general
nature and can be applied to other molecular dynamics simulations.

Before the discussion of the statistical aspect and the improvement of
AMD, we will explain the usual AMD \cite{ONOabc} very briefly for the
convenience of the readers. AMD describes the nuclear many body system
by a Slater determinant of Gaussian wave packets as
\begin{equation}
\Phi(Z)=
\det\Bigl[\exp\Bigl\{-\nu({\vec r}_j - {\vec Z}_i/\sqrt \nu)^2\Bigr\}
\chi_{\alpha_i}(j)\Bigr],
\end{equation}
where the complex variables $Z\equiv\{{\vec Z}_i\}$ are the centers of
the wave packets. We took the width parameter $\nu=0.16\,{\rm
fm}^{-2}$ and the spin isospin states $\chi_{\alpha_i}={\rm
p}\uparrow$, ${\rm p}\downarrow$, ${\rm n}\uparrow$, or ${\rm
n}\downarrow$.  The time evolution of $Z$ is determined by the
time-dependent variational principle and the two-nucleon collision
process. The equation of motion for $Z$ derived from the
time-dependent variational principle is
\begin{equation}
  i\hbar\sum_{j\tau}C_{i\sigma,j\tau}{dZ_{j\tau}\over dt}=
  {\partial{\cal H}\over\partial Z_{i\sigma}^*}.
  \label{eq:AMDEqOfMotion}
\end{equation}
$C_{i\sigma,j\tau}$ with $\sigma,\tau=x,y,z$ is a hermitian matrix,
and $\cal H$ is the expectation value of the Hamiltonian after the
subtraction of the spurious kinetic energy of the zero-point
oscillation of the center-of-masses of fragments,
\begin{equation}
  {\cal H}=\langle H\rangle-{3\hbar^2\nu\over2M}A+T_0(A-N_{\rm F}),
  \label{eq:AMDHamil}
\end{equation}
where $N_{\rm F}$ is the fragment number and $T_0$ is $3\hbar\nu/2M$
in principle, but treated as a free parameter for the adjustment of
the binding energies. In two-nucleon collisions, physical coordinates
${\vec W}_i$ are introduced as
\begin{equation}
  {\vec W}_i=\sum_{j=1}^A \Bigl(\sqrt Q\Bigr)_{ij}{\vec Z}_j,\quad
  Q_{ij} ={\partial \log \langle\Phi(Z)|\Phi(Z)\rangle
           \over \partial({\vec Z}_i^*\cdot{\vec Z}_j)}.
  \label{eq:PhysCoord}
\end{equation}

Except for the two-nucleon collision process, the AMD equation of
motion (\ref{eq:AMDEqOfMotion}) constitutes a kind of Hamilton system,
and therefore one may think that the statistical property of AMD is
not quantum mechanical\cite{OHNISHI-RANDRUP}. It is not true because
AMD can give the exact time evolution of the wave function for the
harmonic oscillator mean field. Namely although the motion of the wave
packet centers is classical, the AMD wave function containing the
spatial and momentum spread of wave packets can have quantum
mechanical statistical property. However, for the process of particle
emission, the statistical property of AMD may be of classical nature,
because the AMD description of the particle emission does not duely
take into account the momentum spread of wave packets. In order to get
physical picture and to enable further development, we calculate the
statistical property of the excited nucleus for the particle emission
process in a new method explained in the following. We added to the
Hamiltonian (\ref{eq:AMDHamil}) the potential wall with a large radius
of the form
\begin{equation}
{k\over2}\sum_i f(|{\vec D}_i-{\vec D}_{\rm CM}|)
\end{equation}
with
\begin{eqnarray*}
&&f(x)=(x-a)^2\theta(x-a),\\
&&\sqrt\nu{\vec D}_i=\mathop{\rm Re}{\vec Z}_i,\quad
{\vec D}_{\rm CM}={1\over A_{\rm tot}}{\textstyle \sum_j} {\vec D}_j,\\
&&a=12\,{\rm fm},\quad k=5\,{\rm MeV/fm^2},
\end{eqnarray*}
and put $A_{\rm tot}$ nucleons into the potential wall and give the
total energy $E_{\rm tot}$. The time evolution is calculated for a
long time ($\sim 10^5$ fm/$c$). As shown in Fig.\
\ref{fig:thermo.vol-no} there are usually a nucleus and nucleons in
the potential wall. Nucleons are sometimes emitted and absorbed by the
nucleus. This situation can be interpreted as the phase equilibrium of
liquid (nucleus) and gas (nucleons). Typical values of the mass number
of the largest fragment $A_{\rm liq}$ and its internal energy $E_{\rm
liq}$ depend on the choice of $A_{\rm tot}$ and $E_{\rm tot}$.  We
select the moments at which $A_{\rm liq}$ takes a given value in order
to get the statistical property of the nucleus of the given mass
number, and the long time average value of $E_{\rm liq}$ and the
temperature are calculated. The temperature $T$, which should be
common to both phases, is calculated as the long time average value of
$\tau$, where $(3/2)\tau$ is the kinetic energy (plus the potential
energy from the wall) per nucleon in gas phase. Since the gas phase is
dilute, the effect of the Pauli principle is neglected. When there are
small fragments in the gas phase, they are excluded in the calculation
of the temperature. The calculated relation between $E_{\rm liq}$ and
$T$ is shown in the left part of Fig.\ \ref{fig:list14b} by squares,
which lie just on the line $\langle E_{\rm liq}\rangle/A_{\rm
liq}=3T-{\rm B.E.}$ of the classical statistics. Each calculated point
corresponds to a specific choice of $A_{\rm tot}$ and $E_{\rm tot}$.
In the calculation of the temperature $T$, i.e., the kinetic energy of
gas nucleons, we only used the central value of the wave packet
$\mathop{\rm Im} {\vec Z}_i$ in momentum space. The neglect of the
momentum width is consistent with the subtraction of the spurious
zero-point kinetic energy from the Hamiltonian (\ref{eq:AMDHamil}).
If we consider the momentum width, the temperature will shift by
$(2/3)T_0$ ($\sim$ 6 MeV) and there will not be any states with
temperature less than $(2/3)T_0$.

\makeatletter\if@floats
\widetext
\begin{figure}
\ifx\epsfbox\undefined\else
  \epsfxsize\textwidth\epsfbox{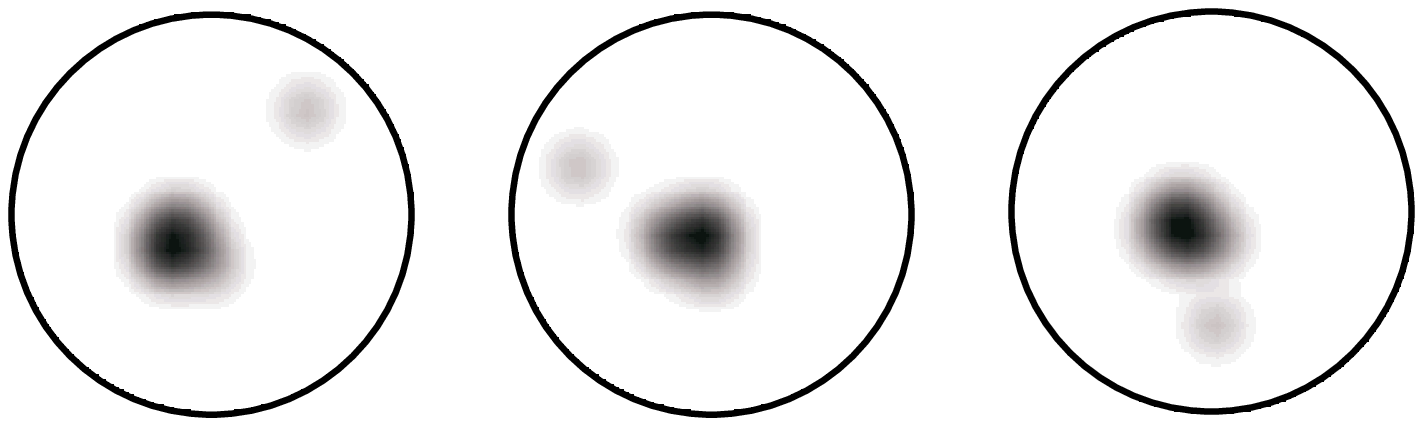}
\fi
\caption{\label{fig:thermo.vol-no}
The density snapshots of the many-nucleon system contained in a
potential wall (shown by circles) calculated with the usual AMD. These
figures are for the total number of the nucleons $A_{\rm tot}=15$ and
the total energy $E_{\rm tot}=-50$ MeV, and the calculated temperature
and the energy of the nucleus are $T=1.48$ MeV and $\langle E_{\rm
liq}\rangle/A_{\rm liq}=-3.65$ MeV for the mass number of the nucleus
$A_{\rm liq}=14$. The Volkov force is used for this calculation.}
\end{figure}
\narrowtext

\widetext
\begin{figure}
\ifx\epsfbox\undefined\else
  \begin{minipage}{0.47\textwidth}
  \epsfxsize\textwidth\epsfbox{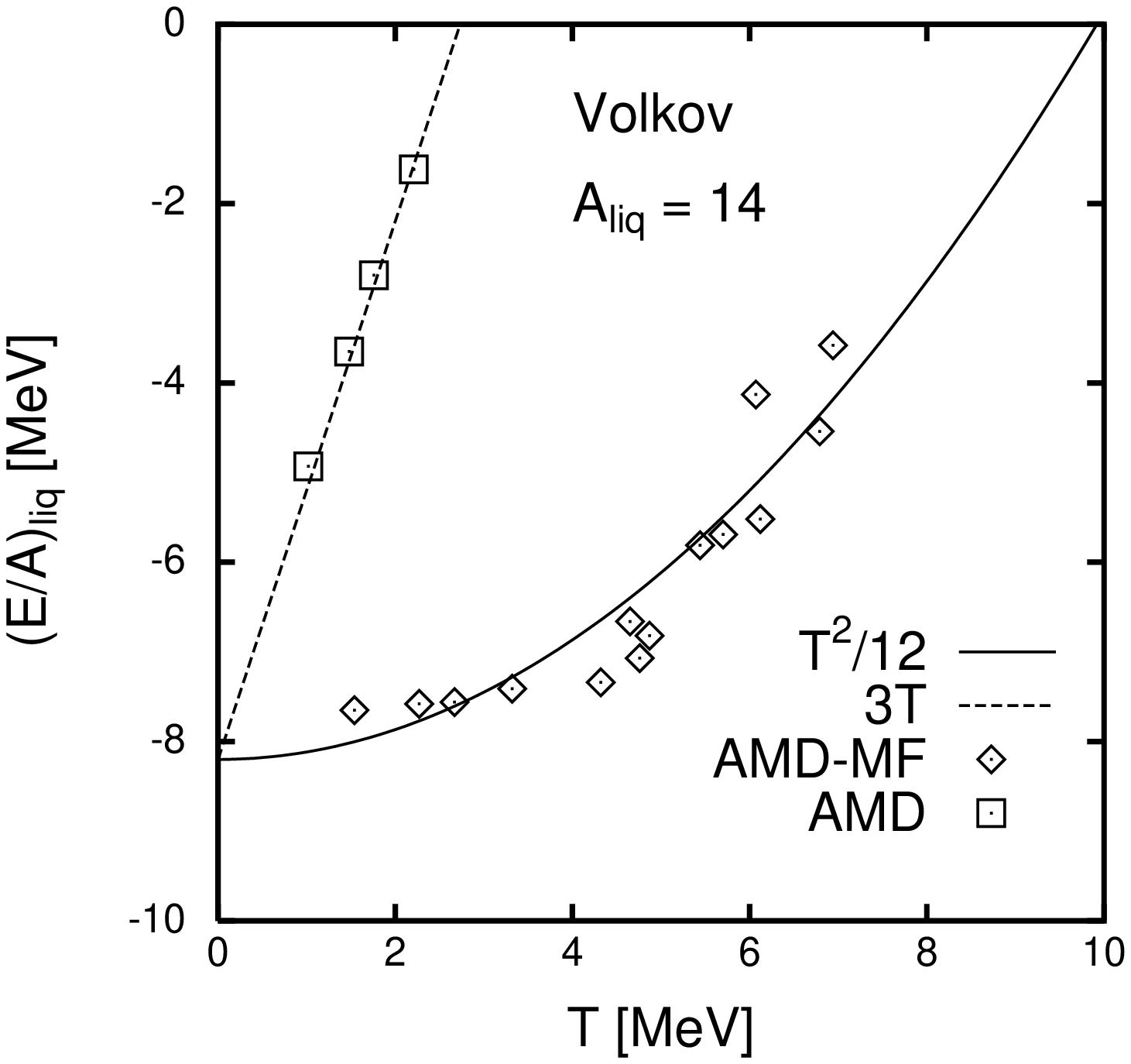}
  \end{minipage}
  \begin{minipage}{0.47\textwidth}
  \epsfxsize\textwidth\epsfbox{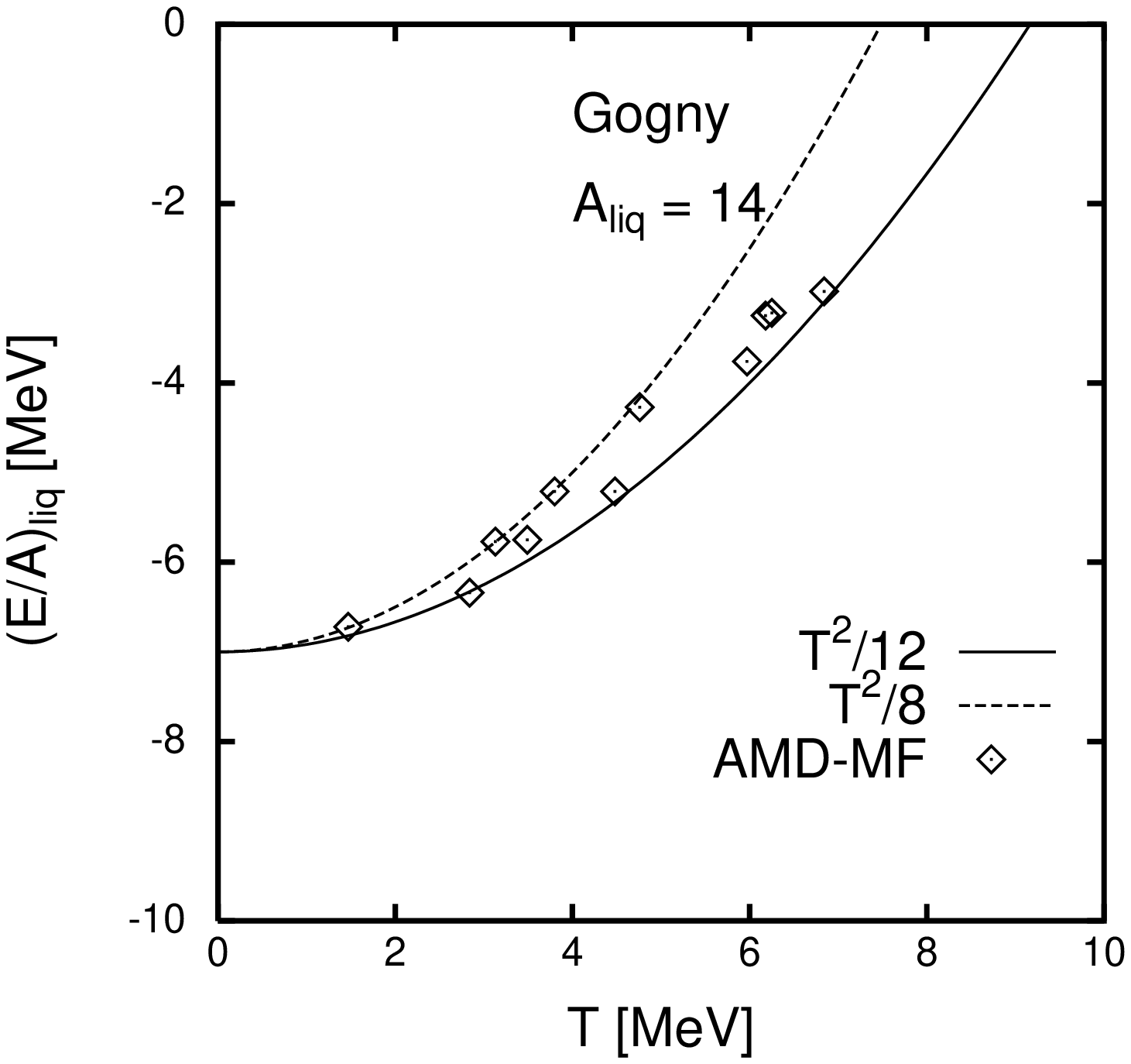}
  \end{minipage}
\fi
\caption{\label{fig:list14b}
The statistical property of the excited nucleus calculated with the
usual AMD and the AMD-MF. The adopted effective interaction is the
Volkov force in the left part, and the Gogny force in the right
part. Lines of $(E/A)_{\rm liq}=3T+{\rm const.}$, $T^2/(12\,{\rm
MeV})+{\rm const.}$, and $T^2/(8\,{\rm MeV})+{\rm const.}$ are drawn
for the comparison.}
\end{figure}
\narrowtext

\widetext
\begin{figure}
\ifx\epsfbox\undefined\else
  \epsfxsize\textwidth\epsfbox{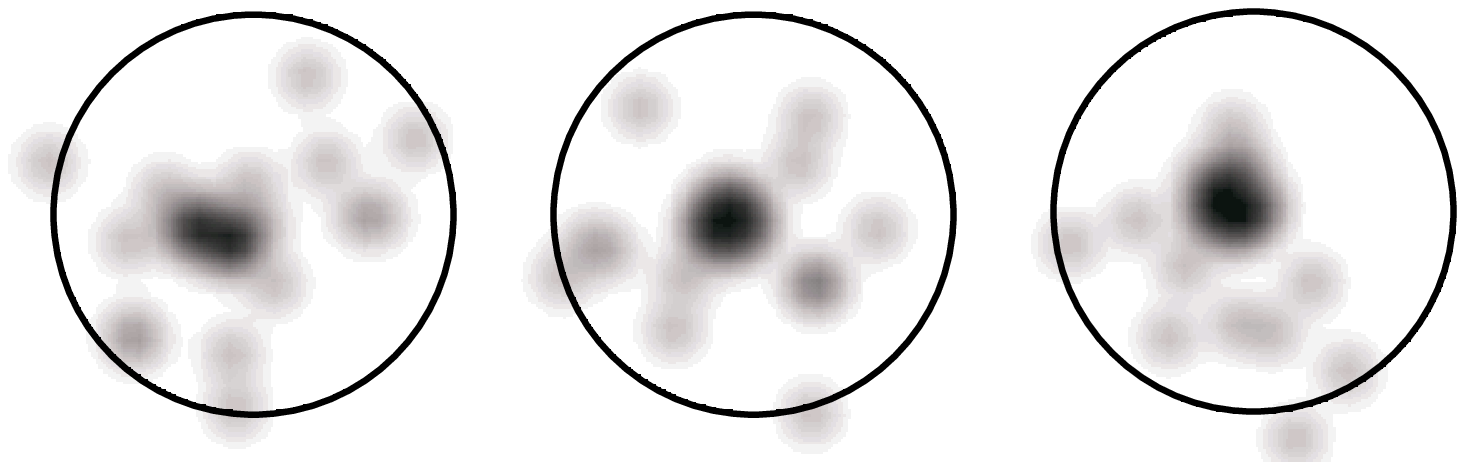}
\fi
\caption{\label{fig:thermo.vol}
Similar to Fig.\ \protect\ref{fig:thermo.vol-no} but calculated with
the AMD-MF. $A_{\rm tot}=28$ and $E_{\rm tot}=60$ MeV for these
figures, and calculated temperature and the energy of the nucleus are
$T=6.12$ MeV and $\langle E_{\rm liq}\rangle/A_{\rm liq}=-5.52$ MeV
for the mass number of the nucleus $A_{\rm liq}=14$.}
\end{figure}
\narrowtext
\fi

Microscopic processes which are important for the liquid gas phase
equilibrium is the emission and the absorption of nucleons by the
nucleus.  The above result indicates that AMD has a problem in the
description of the nucleon emission from the excited nucleus, since
nucleon absorption is naturally described by AMD. Each nucleon in the
nucleus has a momentum width of the wave packet which is an important
part of its Fermi motion in the nucleus. When such a nucleon is
emitted from the nucleus in AMD, it will have the momentum
corresponding to the central value of its momentum distribution in the
nucleus. Especially if the central value is not sufficiently large,
the nucleon will never be able to go out, even if there should be some
probability of going out due to the tail of the momentum distribution.
In order to cure this problem, we regard the emitted nucleon as a
classical particle without momentum distribution or a plane wave,
though it is a wave packet in a nucleus.  And we add a new stochastic
process of giving the momentum fluctuation of the wave packet to a
nucleon when it is being emitted from a nucleus. The total momentum
and energy is conserved by adjusting the state of the nucleus. This
new version of AMD is called AMD-MF and the details will be described
in following paragraphs. By this incorporation of the momentum
fluctuation into AMD, the situation of the liquid-gas phase
equilibrium changes as shown in Fig.\ \ref{fig:thermo.vol}. Even when
the excitation energy of the nucleus is smaller than in the case of
usual AMD (Fig.\ \ref{fig:thermo.vol-no}), the number and the averaged
energy of the gas nucleons are much larger.  The AMD-MF is not a
Hamilton system any more, and the statistical property of the excited
nucleus for the particle emission process has changed drastically into
quantum mechanical one, as shown in Fig.\
\ref{fig:list14b} by diamonds, which is similar to the empirical Fermi
gas formula $\langle E_{\rm liq}\rangle/A_{\rm
liq}=T^2/(12{\rm\,MeV})-{\rm B.E.}$.

It should be noted that our statistical ensemble for an excited
nucleus is between the canonical ensemble and the microcanonical
ensemble. Although both of $E_{\rm liq}$ and $\tau$ are fluctuating in
time, they are correlated due to the energy conservation relation
\begin{equation}
\Delta E_{\rm liq}+{3\over2}A_{\rm gas}\Delta\tau=0
\end{equation}
with $A_{\rm gas}=A_{\rm tot}-A_{\rm liq}$, as long as one can neglect
the probability of the existence of light fragments in gas phases and
the small interaction energy between both phases and among the gas
nucleons. By choosing a large potential wall so that $A_{\rm gas}$ is
very large, it is possible to get the canonical ensemble with small
$\Delta\tau$. On the other hand, if one take small volume for gas
phase, microcanonical ensemble with small $\Delta E_{\rm liq}$ can be
obtained. With our choice of parameters for the case of Fig.\
\ref{fig:thermo.vol}, for example, the energy fluctuation is found to
be $\sqrt{\langle\Delta E_{\rm liq}^2\rangle}/A_{\rm liq}=1.2$ MeV,
which corresponds to the fluctuation of temperature
$\sqrt{\langle\Delta\tau^2\rangle}=0.8$ MeV.

In AMD-MF, the momentum fluctuation is given to a nucleon when it is
being emitted from a nucleus.  At each time step, it is judged whether
each nucleon is in a nucleus or it is isolated in free space. For each
nucleon $i$, a representative point ${\vec r}_i$ is randomly generated
around $\mathop{\rm Re}{\vec W}_i/\sqrt{\nu}$ according to the
Gaussian distribution, where ${\vec W}_i$ is the physical coordinate
(\ref{eq:PhysCoord}). The density at ${\vec r}_i$ without
self-contribution is calculated in an approximate way as
\begin{equation}
\rho_i\sim\biggl({2\nu\over\pi}\biggr)^{3/2}
\sum_{j(\ne i)}
\exp\Bigl[-2(\sqrt\nu{\vec r}_i-\mathop{\rm Re}{\vec W}_j)^2\Bigr].
\end{equation}
If $\rho_i < 0.1\rho_0$ with $\rho_0=0.17$ fm$^{-3}$, the nucleon is
considered to be isolated in free space; else it is in a nucleus. In
the same way, it is checked whether at the previous time step the
representative point ${\vec r}_i-\mathop{\rm Re}({\vec W}_i(t)-{\vec
W}_i(t-\Delta t))$ was in a nucleus or isolated.  Momentum fluctuation
is given to the nucleon $i$ if it was in a nucleus at the previous
time step and it is now isolated in free space.

When it is decided that the nucleon $i$ should be given the momentum
fluctuation now, the following procedures are made.  First of all, the
nucleus from which the nucleon $i$ is being emitted is defined as the
cluster judged by the chain clustering method.  The mass number of the
nucleus is denoted by $A_{\rm nuc}$, which do not include the nucleon
$i$.  Then the relative momentum between the nucleon $i$ and the
nucleus is decided stochastically, by taking account of the fact that
the momentum distribution of the nucleon and the distribution of the
momentum per nucleon of the center-of-mass of the nucleus are
approximately represented by Gaussian distributions
\begin{equation}
\exp\Bigl[-({\vec p}-{\vec P}_i)^2/2\hbar^2\nu\Bigr]
\quad\hbox{and}\quad
\exp\Bigl[-A_{\rm nuc}({\vec p}-{\vec P}_{\rm nuc})^2/2\hbar^2\nu\Bigr],
\end{equation}
respectively,
where
\begin{equation}
{\vec P}_i=2\hbar\sqrt\nu\mathop{\rm Im}{\vec W_i}
\quad\hbox{and}\quad
{\vec P}_{\rm nuc}={2\hbar\sqrt\nu\over A_{\rm nuc}}
                    \sum_{j\in{\rm nuc}}\mathop{\rm Im}{\vec W}_j.
\end{equation}
In order to keep consistency with the method of subtraction of
spurious center-of-mass motion of fragments (\ref{eq:AMDHamil}), the
energy $T_0$ is subtracted from the relative kinetic energy by
reducing the relative momentum.  If the relative momentum is too small
for this procedure, this momentum fluctuation is canceled.

The first candidate of the final state of the momentum fluctuation
process is determined by changing only the relative physical momentum
between the nucleon $i$ and the nucleus to the value decided in the
previous paragraph. However, since the energy is not conserved with
this state, the energy is adjusted to the initial energy with the
least modification of the state by applying the frictional
cooling/heating method with $\lambda+i\mu=\mp i$, for the limited
coordinates ${\vec Z}_j$ with $j$ in the nucleus. And after each step
of the friction, the physical coordinate of the nucleon $i$ and the
physical center-of-mass and the total momentum of the nucleus are
restored to the values of the first candidate.

There are several possibilities for the above procedure to fail.  The
friction leads to a converged energy but still may not reach the
initial energy. Or the transformation from ${\vec W}$ variables to
${\vec Z}$ variables may not exist due to the Pauli-blocking.  In
these cases, the momentum fluctuation is tried again by generating a
different sample of relative momentum.

In summary, a new stochastic process has been added to AMD in order to
improve the description of the nucleon emission process from a hot
nucleus. The momentum fluctuation originating from the momentum width
of the nucleon wave packet in the nucleus is given to the nucleon
stochastically when it is being emitted from the nucleus (AMD-MF).
With this improvement, the quantum mechanical statistical property of
the excited nucleus for the particle emission process has been
obtained, which are calculated in the phase equilibrium of the liquid
phase and the gas phase of nucleons.  It should be noted that we need
not totally modify the dynamics of the AMD equation of motion. As we
mentioned before, we can believe that the AMD wave function gives
fairly good time-evolution of the hot nucleus, because AMD can give
the exact time evolution of the wave function for the harmonic
oscillator mean field. What we have incorporated in this work is a
minor branching process which is brought about by the tail of the
nucleon wave packet. It may be necessary to generalize the
modification so as to improve the cluster emission process in addition
to the nucleon emission. For the application to heavy ion collisions,
other processes caused by the wave packet tail should be considered
such as the nucleon transfer between the projectile and the
target. The wave packet width in spatial coordinate space may also
play some role.  Such extensions have been formulated and calculations
of nuclear collisions have shown that fragmentation processes are
largely influenced by the introduction of the new stochastic process
described here and its extensions.  We will discuss these in other
papers.

\makeatletter\if@floats\else
\widetext
\begin{figure}
\ifx\epsfbox\undefined\else
  \epsfxsize\textwidth\epsfbox{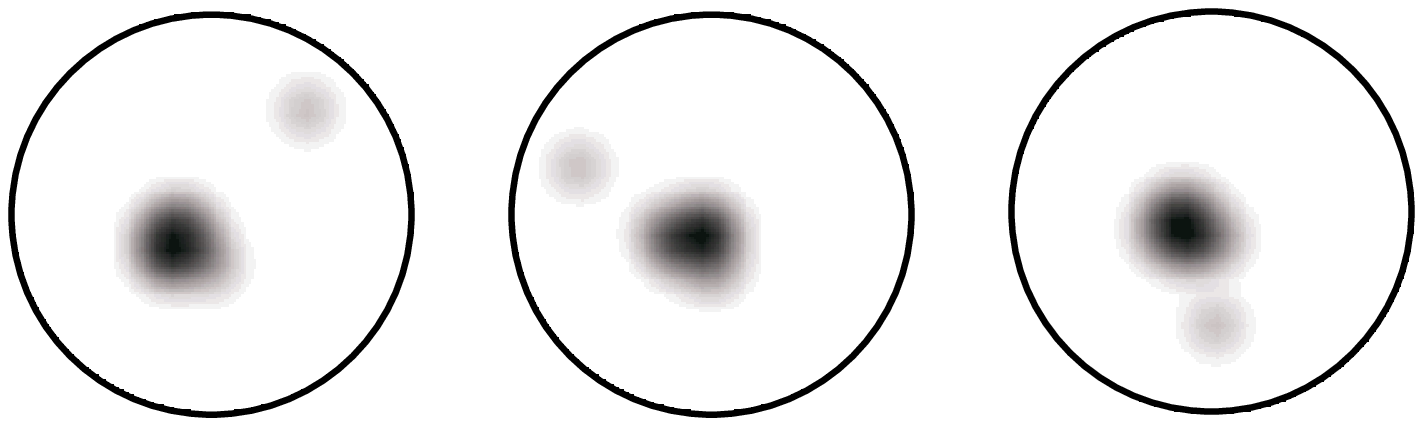}
\fi
\caption{\label{fig:thermo.vol-no}
The density snapshots of the many-nucleon system contained in a
potential wall (shown by circles) calculated with the usual AMD. These
figures are for the total number of the nucleons $A_{\rm tot}=15$ and
the total energy $E_{\rm tot}=-50$ MeV, and the calculated temperature
and the energy of the nucleus are $T=1.48$ MeV and $\langle E_{\rm
liq}\rangle/A_{\rm liq}=-3.65$ MeV for the mass number of the nucleus
$A_{\rm liq}=14$. The Volkov force is used for this calculation.}
\end{figure}
\narrowtext

\widetext
\begin{figure}
\ifx\epsfbox\undefined\else
  \begin{minipage}{0.47\textwidth}
  \epsfxsize\textwidth\epsfbox{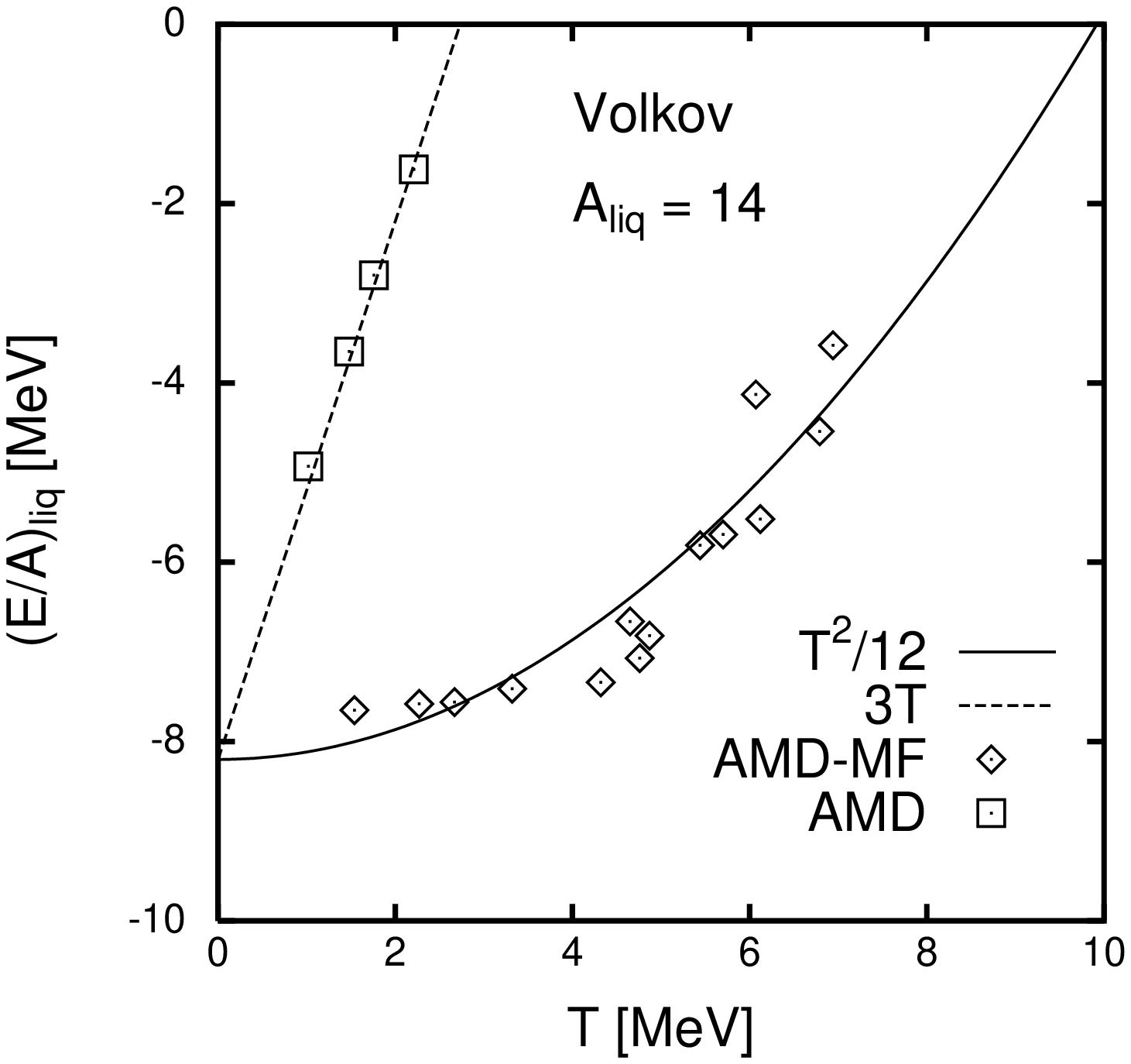}
  \end{minipage}
  \begin{minipage}{0.47\textwidth}
  \epsfxsize\textwidth\epsfbox{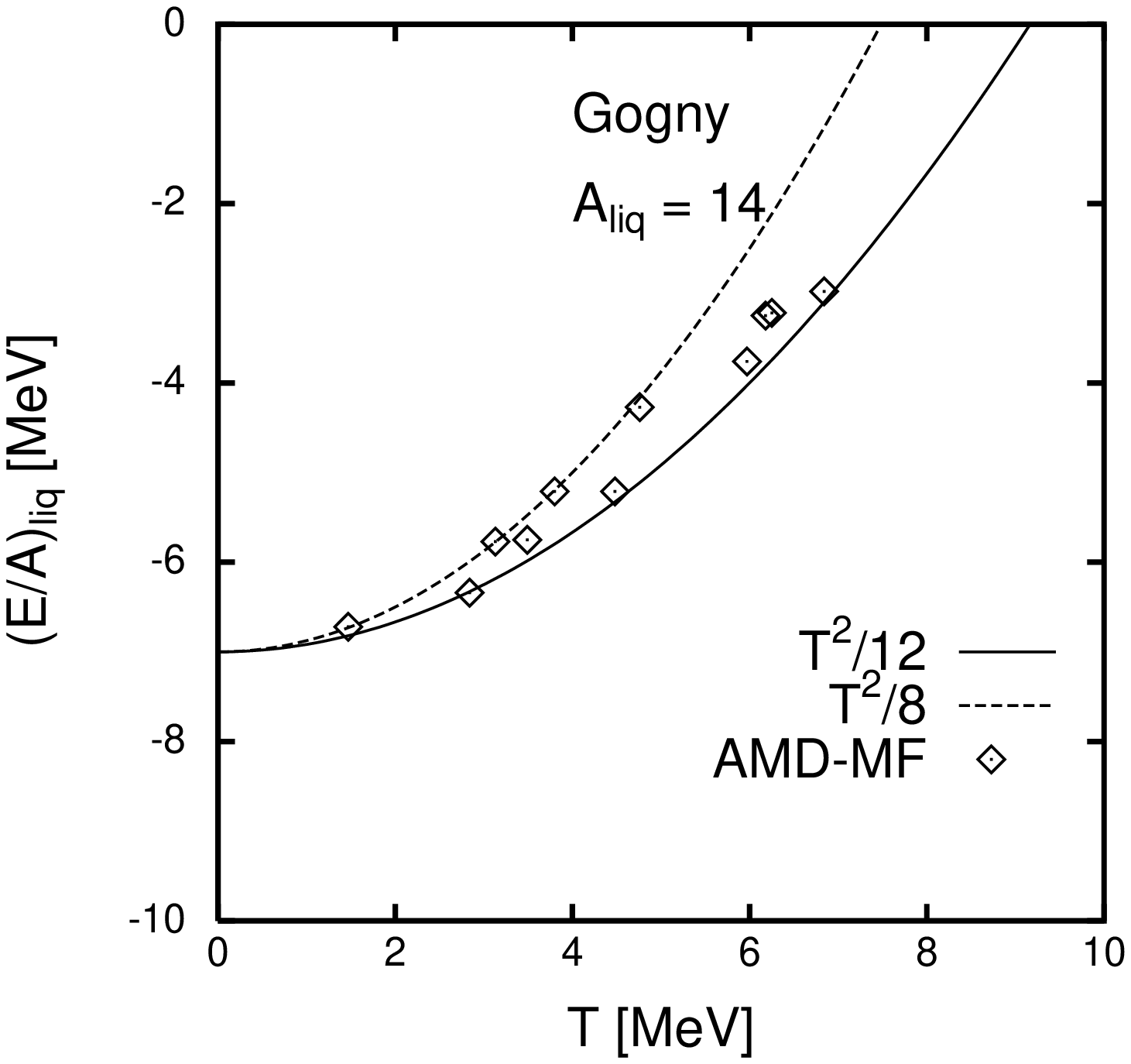}
  \end{minipage}
\fi
\caption{\label{fig:list14b}
The statistical property of the excited nucleus calculated with the
usual AMD and the AMD-MF. The adopted effective interaction is the
Volkov force in the left part, and the Gogny force in the right
part. Lines of $(E/A)_{\rm liq}=3T+{\rm const.}$, $T^2/(12\,{\rm
MeV})+{\rm const.}$, and $T^2/(8\,{\rm MeV})+{\rm const.}$ are drawn
for the comparison.}
\end{figure}
\narrowtext

\widetext
\begin{figure}
\ifx\epsfbox\undefined\else
  \epsfxsize\textwidth\epsfbox{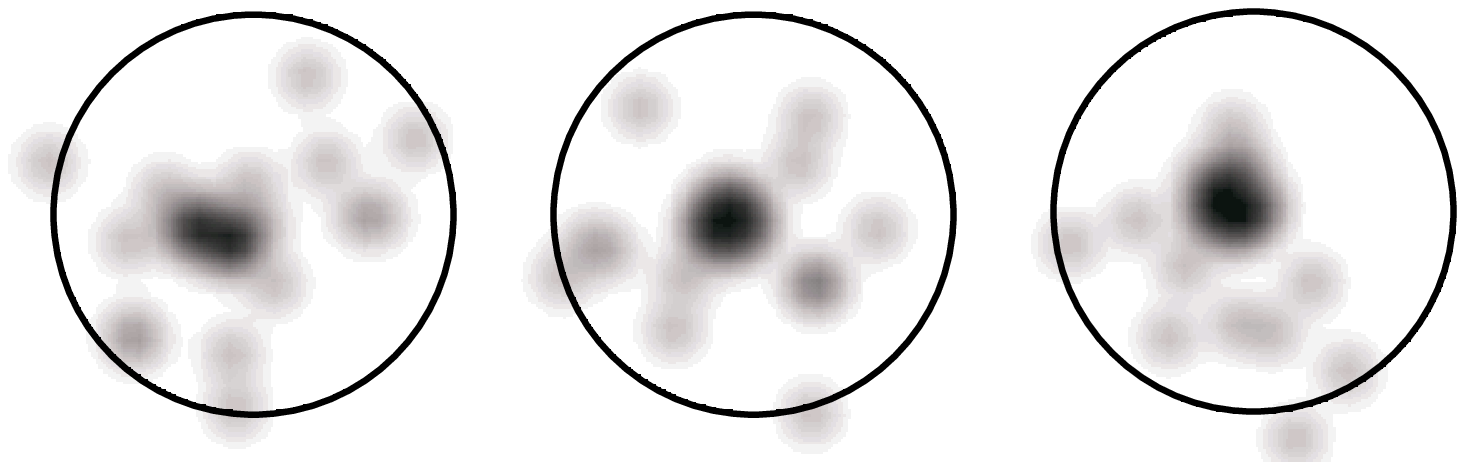}
\fi
\caption{\label{fig:thermo.vol}
Similar to Fig.\ \protect\ref{fig:thermo.vol-no} but calculated with
the AMD-MF. $A_{\rm tot}=28$ and $E_{\rm tot}=60$ MeV for these
figures, and calculated temperature and the energy of the nucleus are
$T=6.12$ MeV and $\langle E_{\rm liq}\rangle/A_{\rm liq}=-5.52$ MeV
for the mass number of the nucleus $A_{\rm liq}=14$.}
\end{figure}
\narrowtext
\fi

\end{document}